\documentclass[conference]{IEEEtran}
\IEEEoverridecommandlockouts
\usepackage[utf8]{inputenc}
\usepackage{amsmath,amssymb,amsfonts}
\usepackage{algorithmic}
\usepackage{graphicx}
\usepackage{balance}
\usepackage[draft]{hyperref}
\usepackage{textcomp}
\usepackage{xcolor}
\usepackage{dblfloatfix}
\usepackage{enumitem}

\def\BibTeX{{\rm B\kern-.05em{\sc i\kern-.025em b}\kern-.08em
		T\kern-.1667em\lower.7ex\hbox{E}\kern-.125emX}}

\begin{document}
	
	\title{Industrial DevOps
		\thanks{This research is funded by the Federal Ministry of Education and Research (BMBF, Germany) in the Titan project (\url{https://www.industrial-devops.org}, contract no.\ 01IS17084B).}
	}

	\author{
		\IEEEauthorblockN{
			Wilhelm Hasselbring\IEEEauthorrefmark{1},
			Sören Henning\IEEEauthorrefmark{1},
			Björn Latte\IEEEauthorrefmark{2},
			Armin Möbius\IEEEauthorrefmark{3},\\
			Thomas Richter\IEEEauthorrefmark{4},
			Stefan Schalk\IEEEauthorrefmark{2}, and 
			Maik Wojcieszak\IEEEauthorrefmark{2}
		}
		\IEEEauthorblockA{
			\IEEEauthorrefmark{1}Software Engineering Group, Kiel University, Department of Computer Science, 24098 Kiel \\
			\IEEEauthorrefmark{2}wobe-systems GmbH, Wittland 2-4, 24109 Kiel, Germany \\
			\IEEEauthorrefmark{3}IBAK Helmut Hunger GmbH \& Co. KG, Wehdenweg 122, 24148 Kiel, Germany\\
			\IEEEauthorrefmark{4}Krause-Biagosch GmbH, Paul-Schwarze-Straße 5, 33649 Bielefeld, Germany\\
			\url{https://www.industrial-devops.org}
		}
	}

	\maketitle

	\begin{abstract}
	
	The visions and ideas of \textit{Industry 4.0} require a profound interconnection of machines, plants, and IT systems in industrial production environments.
	This significantly increases the importance of software, which is coincidentally one of the main obstacles to the introduction of Industry 4.0. 
	Lack of experience and knowledge, high investment and maintenance costs, as well as uncertainty about future developments cause many small and medium-sized enterprises hesitating to adopt Industry 4.0 solutions.
	We propose \textit{Industrial DevOps} as an approach to introduce methods and culture of DevOps into industrial production environments. The fundamental concept of this approach is a continuous process of operation, observation, and development of the entire production environment. This way, all stakeholders, systems, and data can thus be integrated via incremental steps and adjustments can be made quickly.
	Furthermore, we present the Titan software platform accompanied by a role model for integrating production environments with Industrial DevOps. In two initial industrial application scenarios, we address the challenges of energy management and predictive maintenance with the methods, organizational structures, and tools of Industrial DevOps.
		
	\end{abstract}

	\section{Introduction}
	
	The digital transformation of the conventional, manufacturing industry enables a new level of automation in production processes. More and more technical machines and production plants become increasingly intelligent and autonomous. Equipped with network capabilities, they are able to consume and supply data to others.
	This new trend, often referred to as \textit{Industrial Internet of Things} or \textit{Industry~4.0} \cite{Jeschke2017}, confronts the production operator with the challenge of connecting and monitoring the individual machines and devices \cite{Lu2017}. While software used to be only part of the production process, nowadays software increasingly defines the production process itself \cite{Wan2016}, %
	so that the resulting software systems inevitably become more complex.
	In order to ensure a smooth production process these software systems have to be designed with a special focus on reliability, scalability, and adaptability \cite{VogelHeuser2015}. %
	
	As is usual with complex systems, however, this complicates the design, development, and maintenance of such a system. In particular for small and medium-sized enterprises (SMEs), this poses enormous challenges as these SMEs often do not have suitable software development departments. Instead, software is often developed by domain experts (e.g., mechanical engineers) with basic programming skills, but no education in software engineering.
	Alternatives such as establishing a dedicated software development department or delegating software development to specialized external companies are also risky.
	Business or domain requirements and their technical implementations often diverge, so adaptations are cumbersome, costly, and the time until they are released is long.

	Agile and iterative principles, methods, and techniques that are common in other fields of software engineering, such as e-commerce systems \cite{Hasselbring2017, Knoche2019}, can provide solutions to this. Section~\ref{sec:IndustrialDevOps} of this paper describes how these methods can be transferred to the domain of industrial production environments and highlights necessary changes in processes and culture. Section~\ref{sec:Titan} presents a software platform with which these methods can be applied in the manufacturing industry.
	Section~\ref{sec:RoleModel} describes a role model for applying the required organizational and technical structures and Section~\ref{sec:ApplicationScenarios} outlines our two initial application scenarios. Section~\ref{sec:Conclusions} concludes this paper and points out future work.
	
	\section{Industrial DevOps}\label{sec:IndustrialDevOps}

	The traditional separation of software operation from its development leads to several issues due to a lack of communication, collaboration, and integration. DevOps \cite{Bass2015} is a movement to bridge this gap. %
	We propose \textit{Industrial DevOps} as an approach for transferring DevOps values, principles, and methods to industrial systems integration. In this way, we expect that development and operation of such an integrating software system will be improved and the discrepancy between production operators and software developers will be reduced, while ensuring high software quality. The following principles are of crucial importance for Industrial DevOps.
	
	\subsubsection{Continuous Adaption and Improvement Process}
	
	\begin{figure*}[tb]
		\centering%
		\includegraphics[width=0.84\textwidth]{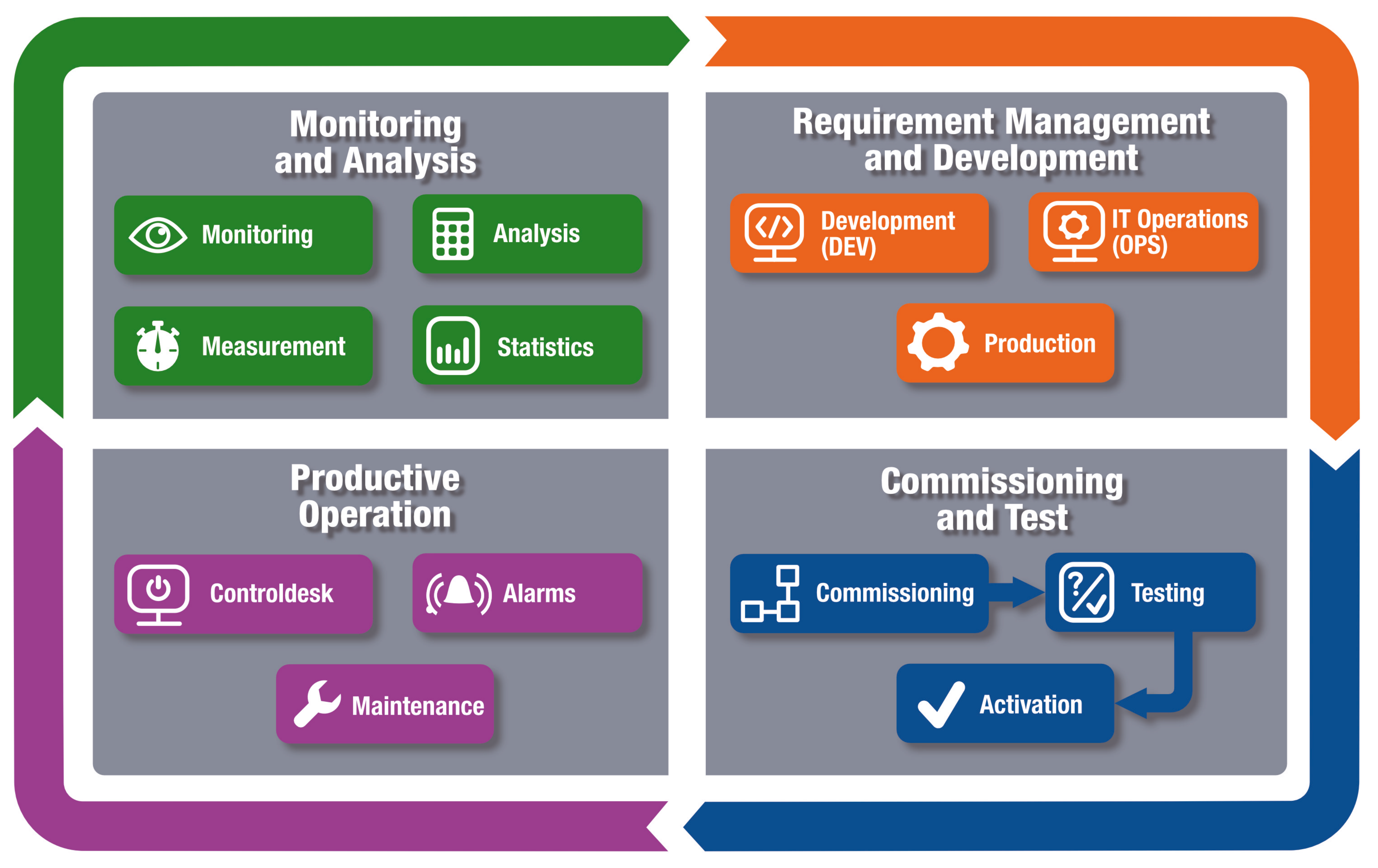}
		\caption{The continuous adaption and improvement process of Industrial DevOps}
		\label{fig:lifecycle}%
	\end{figure*}%
	
	The core element of Industrial DevOps is a coherent, cyclic, and continuous process as illustrated in Fig.~\ref{fig:lifecycle}. During its operation, the software system as well as the production itself is monitored and analyzed. Based on analysis results, new requirements are identified. The implementation of these requirements will either be done by adapting the software or by reconfiguration in the IT operation or production. Adjustments are tested automatically and after passing all tests, the new software replaces the old one in production, where it is monitored again.
	
	Applying this iterative process aims to enable continuous adaptions and improvements of the software. Thus, complex systems are going to be created by starting from simple ones that are extended in incremental steps.
	Another advantage can be found in the maintenance of the systems. While IT systems that run in production after a development phase are provided with updates for a certain period of time, followed by a certain period of service, the actual operating time is often considerably longer. As a result, operation becomes increasingly complex and risky, but at the same time migrating to another system becomes more and more difficult. The cyclic process of Industrial DevOps is intended to achieve that development, updates, and service become continuous processes that run permanently parallel to the operation of the system.
	
	\subsubsection{Lean Organizational Structure}
	
	When applying external software solutions, this often requires enterprises to adapt their production processes to these solutions. %
	Industrial DevOps comes with an organizational structure to prevent this and instead support the alignment of the software with the production process.
	
	A key factor of Industrial DevOps is that requirements are discussed between all stakeholders and people from different business units are brought together. This practice is referred to as  BizDevOps as a more general term \cite{Gruhn2015}. Industrial DevOps extends this by applying BizDevOps to integrating systems and data of industrial production environments. For the implementation of these requirements, domain experts and developers work closely together. Moreover, the development provides appropriate means, which allow domain experts to solve the domain problems themselves to a certain extent. This is accomplished by an extendable software platform, which is owned by the user. Ownership can be direct or indirect if the software is owned by an open source community.
	
	Information from all levels is available for anyone in the value stream through monitoring. Even if the information that is required differs, the source of information is consistent and allows investigation about cause and effect (culture of causality). This should enable lean organizational learning \cite{Knuf2000}.

	\subsubsection{Customer-Centric Value Generation}
	
	Organizations who apply customer-centric value generation align all departments to contribute to this goal \cite{Denning2018}. All activities in the organization which do not directly or indirectly contribute to this goal are abandoned. Some activities which are required for organizational or compliance reasons are organized to stay out of the way of the overall goal.
	Instead of wasting huge amounts of money in market research, experiments are started in a way that new features, functions, or products are offered to customers and direct measurements are made by utilizing IT to find out the customers response. These measurements are not only done once, but continuously.

	\begin{figure*}[tb]
		\centering%
		\includegraphics[width=0.84\textwidth]{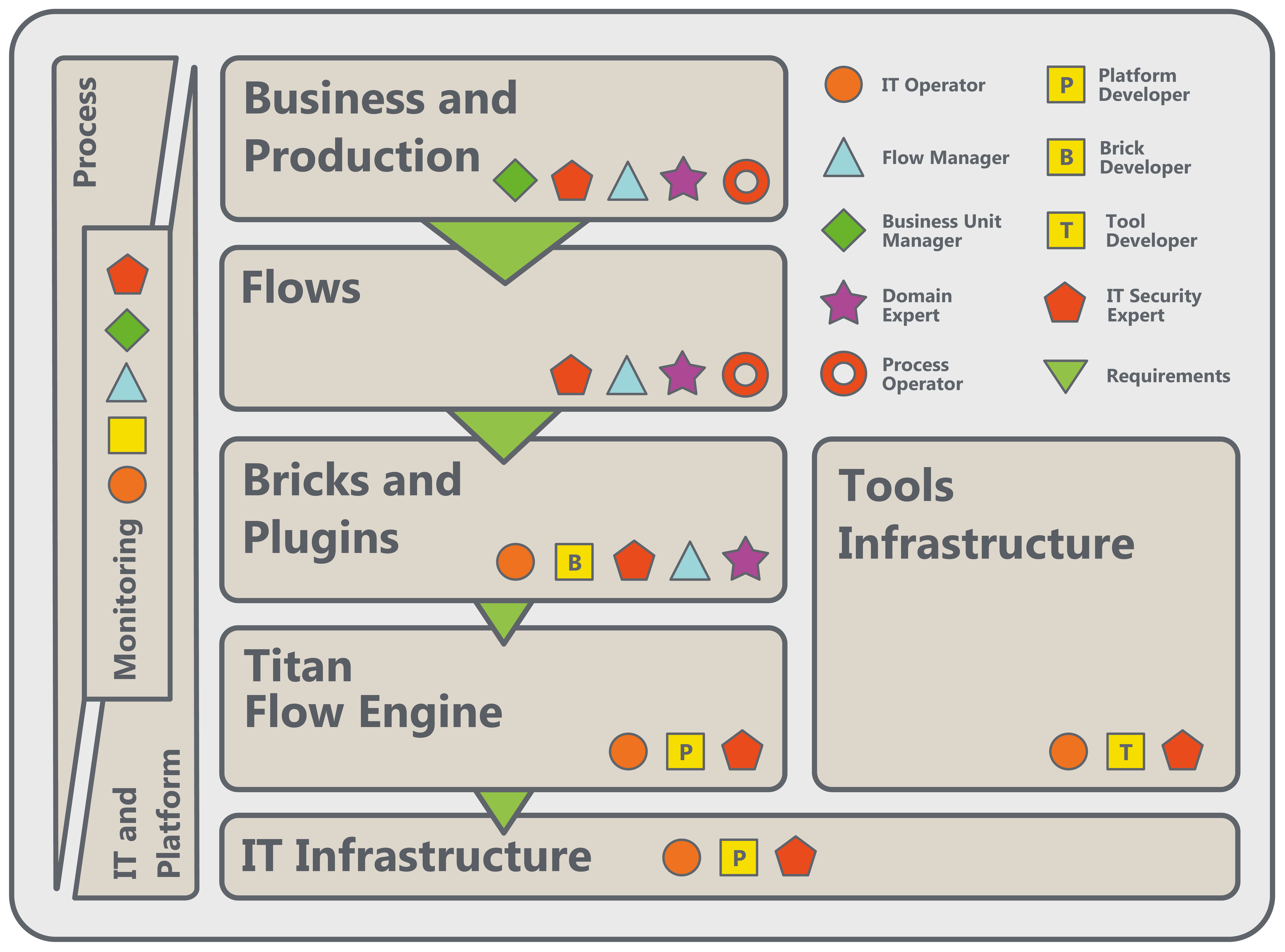}
		\caption{The Titan role model}
		\label{fig:role-model}%
	\end{figure*}%
	
	\section{The Titan Platform}\label{sec:Titan}
	
	Titan is a software platform for integrating and monitoring industrial production environments using Industrial DevOps. In order to integrate tools or infrastructure (e.g., machines, IoT %
	devices, software, or databases), modular software components are created. These components, in Titan called \textit{bricks}, serve for integrating exactly one such system. %
	This allows on the one hand to exchange the underlying tool or infrastructure easily, but on the other hand also to replace the integrating software implementation.
	To connect bricks (and thus the underlying systems of the production), Titan applies the principles of flow-based programming \cite{Morrison2010} and calls a network of connected bricks a \textit{flow}.
	
	Titan provides a graphical modeling language, which enables the organization to graphically model the integration and, based on this, to configure the integrated system. Hence, there is no programmer required to perform these configurations or (pre-configured) changes. Instead, a person that has received training on the modeling language can work together with a domain expert to do this.
	
	Thus, this platform allows to react efficiently and effectively to changing requirements in production. Furthermore, it enables conducting experiments with little effort and to adjust quickly in case of experiment failures. Titan integrates the required monitoring, which includes resource consumption, system utilization, and business data.

	\section{The Titan Role Model}\label{sec:RoleModel}

	Industrial DevOps involves people from different domains, business units, and organizations. The Titan role model (see Fig.~\ref{fig:role-model}) defines roles to be assigned to people or groups that use, operate, or develop the Titan platform and describes their responsibilities as well as how they interact.

	A \textbf{Developer} is a trained software engineer. This person has learned to develop software in many programming languages. Many different aspects of software development like code quality or version management are covered. In Industrial DevOps, many developers who are spread over many teams in different organizations work together on a single system. The role model introduces the following sub-roles:
	The \textbf{Platform Developer} works on new features and improvements of the Titan platform itself.
	The \textbf{Tool Developer} creates software tools or interfaces for hardware that is integrated into the system.
	The \textbf{Brick Developer} creates bricks and works closely together with the Tool developers.

	The \textbf{IT Operator} is trained in installing, configuring, and running IT infrastructure. This person is also responsible for installing, updating, configuring, and running software applications.
	
	The \textbf{Flow Manager} is a person that is closely related to the business, but with a special training to master all aspects of flow modeling. This person works as a connection between business and technology. The Flow Manager implements new flows and flow changes according to business requirements.
	
	The \textbf{Business Unit Manager} can be located on all levels of the hierarchy. This person knows about the business requirements and works together with the Flow Manager, the Domain Expert, and the Process Operator to refine requirements.
	
	The \textbf{Domain Expert} owns deep knowledge about the domain at hand.
	
	The \textbf{Process Operator} benefits from a new flow or a change. This person is involved in utilizing the flow in the actual production environment.
	
	The \textbf{IT Security Expert} is responsible for the IT security domain of the whole organization. The IT Security Expert works together with development and operations through all levels of the system.
	
	As an extension to regular DevOps roles, the Titan role model introduces business roles, domain experts, flow managers, and process operators, which do not need to be developers. %

	\section{Initial Industrial Application Scenarios}\label{sec:ApplicationScenarios}

	Enterprises, in particular in the manufacturing industry, face the challenge of reducing and optimizing their energy consumption for economic and ecological reasons. This requires in-depth monitoring and analysis of the individual energy consuming devices, machines, and plants. Doing this,	also referred to as energy management, serves as an initial application scenario for Industrial DevOps and Titan. We designed a monitoring infrastructure, which monitors, analyses, and visualizes the electrical power consumption in industrial production environments \cite{Henning2019}. Employing the Titan platform, it is able to integrate different kinds of sensors that use different data schemata, formats, and protocols.  Referring to the Titan role model, this monitoring infrastructure enables Business Unit Managers, Domain Experts, and Process Operators to gain detailed insights into the energy usage and to identify saving potentials. Thus, it is possible to perform and evaluate optimization measures across departmental boundaries within the company.
	
	Predictive maintenance is another initial application scenario we address with Industrial DevOps and Titan. Titan can serve as a platform, which detects faults in machines and production plants in advance on the basis of data from integrated sensors and systems. An essential aspect here is that not only data from a single machine and production plan is considered for analysis. Instead, measurements and events from of the entire production as well as environmental influences can be taken into account. The predictive maintenance function thus has a much broader amount of data available and can therefore also include non-obvious influences. With the insights gained from the data, maintenance intervals can be optimized. Additionally by returning analysis results to the machines, they can optimize themselves.
			
	\section{Conclusions and Future Work}\label{sec:Conclusions}

	We propose Industrial DevOps as an approach to transfer practices, methods, and culture of DevOps for integrating systems in industrial production environments. It suggests an organizational structure in which domain experts, business managers, developers, and operators work together on solutions. Production environments should be integrated, configured, and operated in a validated learning loop to manage changes and to control costs.

	Future work lies in evaluating the proposed principles, methods, and organizational structures of Industrial DevOps in different industrial application scenarios. Therefore, we further develop the Titan platform. To overcome the disadvantages of proprietary integration systems, we will release it as free and open source software. %
	We are already using \textit{Clean Code} approaches in the development to prevent software quality from degrading over time \cite{Latte2019}.
	Along with including quantitative quality characteristics such as performance in the DevOps cycle \cite{Waller2015} as well as rigorous monitoring in operation, this is supposed to contribute in creating a long-living software platform \cite{Goltz2015}.

	\bibliographystyle{myIEEEtran}
	\bibliography{IEEEabrv,references}

\end{document}